\documentclass[aps,prl,nofootinbib,floatfix,twocolumn]{revtex4-2}
\usepackage{amsmath,amssymb,color,graphicx,bm,hyperref,mathrsfs}
\usepackage{physics}

\usepackage{multibib}

\definecolor{red}{rgb}{0.8,0,0}
\definecolor{RED}{rgb}{0.8,0,0}
\definecolor{violet}{rgb}{0.4,0,0.4}
\definecolor{green}{rgb}{0,0.5,0.0}
\definecolor{GREEN}{rgb}{0,0.5,0.0}
\definecolor{navy}{rgb}{0.0,0.0,0.6}
\definecolor{orange}{rgb}{0.8,0.2,0.0}
\definecolor{blue}{rgb}{0.3,0.0,0.8}

\begin{document}
\title{\textbf{Spontaneous entanglement leakage of two static entangled Unruh-DeWitt detectors}}
\author{Dipankar Barman}
\email{dipankar1998@iitg.ac.in}
\author{Angshuman Choudhury}
\email{c.angshuman@iitg.ac.in}  
\author{Bhushan Kad}
\email{k.bhushan@iitg.ac.in}
\author{Bibhas Ranjan Majhi}
\email{bibhas.majhi@iitg.ac.in}
\affiliation{Department of Physics, Indian Institute of Technology Guwahati, Guwahati 781039, Assam, India.}

\begin{abstract}
Two entangled two-level Unruh-DeWitt detectors, which are in rest, spontaneously loose entanglement when at least any one of them is not isolated from the environment quantum fields. For eternal interaction between the detectors and environment, the spontaneous emission from the detectors' exited states and vacuum fluctuations of field influence this negative effect. Consequently, it suggests that two entangled qubits become less communicated during their free-fall towards the black hole horizon.
\end{abstract}
\maketitle

{\it{Introduction.}}--
Quantum entanglement, known as a resource of the quantum information, is key to many quantum technologies such as quantum cryptography \cite{PhysRevLett.67.661,Yin:2020aa}, quantum teleportation \cite{Hotta:2008uk, Hotta:2009, Frey:2014, Matson:2012aa}, quantum computation \cite{Mooney:2019aa}, {\it etc}. Such phenomena drew lot of attention over the past few decades both in non-relativistic (NR) and relativistic regimes. In NR quantum systems, lots of studies have been carried out to understand it \cite{PhysRev.47.777, PhysicsPhysiqueFizika.1.195, PhysRevLett.23.880}, including its realisation in the lab \cite{PhysRevLett.28.938, PhysRevLett.49.1804, PhysRevLett.80.3891, Hensen:2015ccp,PhysRevLett.118.060401}. Moreover this phenomena became very important to understand the quantum nature of gravity \cite{PhysRevLett.119.240401,PhysRevLett.119.240402}, the black hole information paradox \cite{hawking1975, Almheiri:2012rt, Marolf_2017}, black hole thermodynamics \cite{Brustein:2005vx,Solodukhin:2011gn}, {\it etc}.

Study of quantum entanglement in the relativistic framework can provide a broader perspective towards the reality. Interestingly the existence of entanglement in vacuum state of a quantum field is capable of harvest entanglement between a pair of two-level detectors (known as Unruh-DeWitt (UD) detectors), even if they are causally disconnected \cite{VALENTINI1991321,Reznik:2002fz,SUMMERS1985257, doi:10.1063/1.527733, Reznik:2003mnx,Thomas:2021aa}. This process of swapping field entanglement to detectors is sensitive to the type of motion of UD detectors \cite{FuentesSchuller:2004xp, Martin-Martinez:2015qwa,Salton:2014jaa, Koga:2018the, Koga:2019fqh}, the switching function of the interactions \cite{Pozas-Kerstjens:2015gta}, the nature of the background fields \cite{ Kukita:2017etu, Barman:2021bbw, Perche:2021clp}, 
presence of black holes \cite{Henderson:2017yuv, Tjoa:2020eqh, Cong:2020nec, Gallock-Yoshimura:2021yok, Barman:2021kwg} and other curved spacetimes \cite{VerSteeg:2007xs, Henderson:2018lcy,Barman:2021kwg}, which have been very active areas for last few years. Moreover, the quantum entanglement phenomenon appears to be frame dependent -- the measure of entanglement changes as one describes with respect to other reference frames \cite{PhysRevA.55.72, PhysRevLett.88.230402, PhysRevLett.94.078901, PhysRevLett.89.270402, PhysRevLett.91.180404, FuentesSchuller:2004xp, Chowdhury:2021ieg}. In a bipartite system the measure of entanglement is fruitfully quantified by {\it Negativity} as well as {\it Concurrence} \cite{Peres:1996dw, Hill:1997pfa,Wootters:1997id}.

In this letter we intend to address a fundamentally important question -- whether entanglement between systems remain intact when they are not isolated from the environment? In our environment the background fields always contain vacuum fluctuation energy and as the entanglement in vacuum state of the background fields swaps to a pair of UD detectors (see e.g.  \cite{Reznik:2003mnx, Martin-Martinez:2015qwa, VerSteeg:2007xs, Kukita:2017etu, Cong:2020nec, Henderson:2018lcy,Barman:2021bbw, Henderson:2017yuv, Tjoa:2020eqh, Gallock-Yoshimura:2021yok, Barman:2021kwg, Pozas-Kerstjens:2015gta,Salton:2014jaa, Koga:2018the, Koga:2019fqh, Barman:2021bbw, Perche:2021clp})), it is natural to investigate whether the swapping of vacuum entanglement has any influence in entangled systems. To get the answer to this we consider a relativistic model where two initially entangled UD detectors are individually interacting with the background fields. For simplicity the fields are chosen to be real scalar ones and the interaction is monopole type. In order to avoid the effect of motion we consider both the detectors to be static eternally with respect to the lab frame in Minkowski spacetime. Also to avoid complexity of calculation and perform an analytical analysis, we choose eternal interaction between the field and detectors. The investigation, like earlier various analysis, is done till second order in perturbation series.

We observe that even if the detectors are eternally static with respect to the lab frame, they lose entanglement communication while interacting with the environment (here the background scaler field). This feature is a bit unexpected as previous results are in favour of entanglement harvesting due to entanglement swapping (like mentioned in \cite{Reznik:2003mnx, Martin-Martinez:2015qwa, VerSteeg:2007xs, Kukita:2017etu, Cong:2020nec, Henderson:2018lcy,Barman:2021bbw, Henderson:2017yuv, Tjoa:2020eqh, Gallock-Yoshimura:2021yok, Barman:2021kwg, Pozas-Kerstjens:2015gta,Salton:2014jaa, Koga:2018the, Koga:2019fqh, Barman:2021bbw, Perche:2021clp}). Therefore, if two entangled qubit are left open in the environment, they will lose entanglement. We find that such leakage of entanglement within this simple model is caused by collective effects of spontaneous emission of the individual detector and vacuum fluctuation of quantum field. Moreover, we argue that the leakage is unavoidable even for other type of switching function related to interaction. In the latter situation other effects (like excitation of detectors, {\it etc}.) may also contribute. Interestingly, if any one of the detectors is kept shielded from the environment (while the other one is open to environment quantum field) then also the composite system suffers from entanglement degradation. But in the latter situation the same will be less than the situation where both the detectors are switched on.
 
Let us now proceed towards the calculation to justify our claim.

%%%%%%%%%%%%%%%%%%%%%%%%%%%%%%%%%%%%%%%%%%
{\it The UD detector model.}-- 
Consider a pair of UD detectors, $A$ and $B$, with energy gap $\Delta{E}_{j} = E_{e_j}-E_{g_j}~(j=A,\,B)$. The detectors are initially entangled and the initial quantum state is taken to be 
\begin{equation}
|D\rangle=\alpha|g_{A}g_{B}\rangle+\gamma|e_{A}e_{B}\rangle~,
\label{B4}
\end{equation} 
with $\alpha$ and $\gamma$ are chosen to be real and satisfy $\alpha^2+\gamma^2=1$. Here $|g_{j}\rangle$ and $|e_{j}\rangle$ are the ground and exited states of $j^{th}$ detector, respectively and $|g_{A}g_{B}\rangle \equiv |g_{A}\rangle \otimes |g_{B}\rangle$, so on.
The detectors are at rest in ($3+1$)-dimensional Minkowski spacetime and hence their trajectories are denoted as
\begin{eqnarray}
\label{trajectories}
&&t_{A}=\tau_{A},~\mathbf{x}_A=0~;
\nonumber
\\
&&t_{B}=\tau_{B},~\mathbf{x}_B=\mathbf{d}~,
\end{eqnarray}
where $\mathbf{d}$ is a constant vector (measures the distance between the detectors) and $\tau_j$ is proper time of $j^{th}$ detector.
The action for the interaction between detector and background quantum field $\phi(x)$ is taken to be monopole type \cite{book:Birrell}:
\begin{equation}
S_{int}=\sum_{j=A,B}C_{j}\int{d\tau_{j}} \chi_j m_{j}(\tau_{j})\phi(x_{j}(\tau_j))~,
\end{equation}
where $C_{j}$ is the coupling constant of the interaction and $m_{j}(\tau_{j}) = e^{iH_{j}\tau_{j}}(|g_{j}\rangle\langle e_{j}|+|e_{j}\rangle\langle g_{j}|)e^{-iH_{j}\tau_{j}}$, is the monopole operator of the $j^{th}$ detector, whose free Hamiltonian is given by $H_{j}$. Here $\chi_j$ is the switching function which determines the duration of interaction.
The initial state of the composite system (field and detectors system) is considered to be $|in\rangle=|0_{M}\rangle\otimes|D\rangle$, where $|0_{M}\rangle$ denotes the vacuum state of field in Minkowski spacetime.

The initial density matrix of the detectors' system (by tracing out the field degrees of freedom) is given by
\begin{equation}
\rho_{AB}(t_{in})=\begin{pmatrix}
\alpha^{2}&0&0&\alpha\gamma\\0&0&0&0\\0&0&0&0\\\gamma\alpha&0&0&\gamma^{2}
\end{pmatrix}\,.
\label{B2}
\end{equation}
The later time this density matrix is determined as $\rho_{AB}(t) =\text{Tr}_{\phi}(\text{T}e^{iS_{int}}|in\rangle\langle in| \text{T}e^{-iS_{int}})$, where $\text{T}$ denotes the time ordering. Till the second order in perturbation series this turns out to be \cite{Chowdhury:2021ieg}
\begin{equation}
\label{LTDM}
\rho_{AB}(t)
=\begin{pmatrix}
a_{1}&0&0&a_{2}\\
0&b_{1}&b_{2}&0\\
0&c_{1}&c_{2}&0\\
d_{1}&0&0&d_{2}\\
\end{pmatrix}~,
\end{equation}
with $t>t_{in}$.
The explicit forms the matrix elements are presented in \cite{Chowdhury:2021ieg} (see also Supplemental Material). Consider the detectors are identical and so $\Delta E_A = \Delta E_B \equiv \Delta E$. Also for simplicity assume the coupling constants $C_A, C_B$ are same; i.e. $C_A=C_B\equiv C$. For the present model with $\chi_j=1$ then they reduce to
\begin{equation}
\label{rhoABelements}
\begin{aligned}
a_1 &=
\gamma \gamma(1- C^{2}P_A^{\prime \prime}- C^{2}P_B^{\prime \prime})\,,\\
a_2 &=
\gamma \alpha(1-C^{2}M_A-C^{2} M_B) \,,
\\
b_1 &=
\gamma \gamma C^{2}P_B^{\prime \prime}\,,\\
b_2 &=\gamma \gamma C^2X_{A B}^{\star}\,,\\
c_1 &= \gamma \gamma C^2X_{A B}\,,\\
c_2 &=\gamma \gamma C^{2}P_A^{\prime \prime}\,,\\
d_1 &=\alpha \gamma(1-C^{2}M_A^{\star}-C^{2}M_B^{\star})\,,\\
d_2 &=\alpha \alpha\,;
\end{aligned}
\end{equation}
where 
\begin{eqnarray}
&&P_{j}''(\Delta{E})=\int\int d\tau_{j}d\tau_{j}'e^{-i\Delta{E}(\tau_{j}-\tau_{j}')}G_{W}(x_{j}',x_{j}) 
\nonumber
\\
&&= P_j(-\Delta E)~;
\nonumber
\\
&&M_{j}(\Delta{E})=\int\int d\tau_{j}d\tau_{j}'e^{i\Delta{E}(\tau_{j}-\tau_{j}')}\theta(\tau_{j}-\tau_{j}')
\nonumber
\\
&&\times(G_{W}(x_{j}',x_{j})+G_{W}(x_{j},x_{j}'))~;
\nonumber
\\
&&X_{AB}(\Delta{E})=\int\int d\tau_{A}d\tau_{B}'e^{i\Delta{E}(\tau_{B}'-\tau_{A})}G_{W}(x_{B}',x_{A}).
\label{B1}
\end{eqnarray}
In the above $G_{W}(x_{j}',x_{j})$ is the positive frequency Wightman function. As $P_j(\Delta E)$ denotes the transition probability to exited state of $j^{th}$ detector (see e.g. \cite{book:Birrell}), the first term in (\ref{B1}) corresponds to the de-excitation of it. Since the detectors are static, therefore $P_{j}''$ does not have any contribution due to the relative motion of the detectors. Therefore it is completely given by the {\it spontaneous emission probability} \cite{Akhmedov:2007xu}. On the other hand, $G_{W}(x_{j}',x_{j})+G_{W}(x_{j},x_{j}')$ in the second term can be realised as $\bra{0_M}\{\phi(x_j'),\phi(x_j)\}\ket{0_M}$. Therefore $M_j$ is determined from the anti-commutator of the scalar field and as the expectation value of anti-commutator depends on the field state under consideration (contrary to the commutator of field, whose expectation value is independent of state), $M_j$ arises purely due to the {\it vacuum fluctuation} of field.

Before going to evaluate the integration, let us mention about the choice of $\chi_j$. In detector transition-related studies, one can start with an adiabatic interaction switching function $\chi_{j}(\tau_{j})=e^{-s|\tau_{j}|}$ to build up a model (e.g see \cite{Takagi01031986,book:Birrell}). This helps to suppress spurious transient effects. In that case the interaction is kept switched on for a duration $\tau_j\sim s^{-1}$ with the restriction $s<< \Delta E$. Moreover to fulfil the adiabatic condition, the interaction is switched on and off in an infinitely slow process. Then if the interaction starts at $t_{in} = \tau_{j}^{in}=-\tau_0$ and ended at $t=\tau_{j}^{f} = \tau_0$, one must take the limits $\tau_0\to \infty$ and $s\to 0$. In that case the limits of integrations are from $-\infty$ to $\infty$ and $\chi_j$ can be chosen to unity. Using this sprit we have chosen $\chi_j=1$ in the present analysis so that the interaction starts at the asymptotic infinite-past (same has also been done in \cite{book:Birrell}). 
For that, the detectors and the field are to be in a product state at the initial time, i.e., $t_{in}=-\infty$.

Evaluation of the integrations yield
\begin{eqnarray}\label{Expressions}
&&P_{j}''=\frac{\delta(0)}{2c^{3}}\sqrt{\Delta{E}^{2}-m^{2}c^{4}}\equiv P''~;
\label{Pj''}
\nonumber
\\
&&\text{Re}(M_{j})=\frac{\delta(0)}{4c^{3}}\sqrt{\Delta{E}^{2}-m^{2}c^{4}}\equiv M~;
\label{Mj}
\nonumber
\\
&&X_{AB}=
\frac{\delta(0)}{2c^{3}}\sqrt{\Delta{E}^{2}-m^{2}c^{4}}~\frac{\sin\left(\frac{d}{c}\sqrt{\Delta{E}^{2}-m^{2}c^{4}}\right)}{\left(\frac{d}{c}\sqrt{\Delta{E}^{2}-m^{2}c^{4}}\right)},
\label{XAB}
\end{eqnarray}
where ``$\textrm{Re}$'' denotes the real part only. In the above $c$ is the velocity of light in the free space.
Since both of the detectors are static with respect to the lab frame on Minkowski spacetime, we have $P_{A}''=P_{B}''$ and $M_{A}=M_{B}$. The Dirac-delta functions in these expressions are arising due to infinite time integrations (i.e. $\chi_j(\tau_j)$ has been chosen to be unity). Then to give a meaning one can interpret the quantities divided by the delta functions as a rate of these quantities by considering 
\begin{equation}
\delta(0) = \lim_{T\to\infty}\frac{1}{2\pi}\int_{-T/2}^{T/2} du~,
\label{B6}
\end{equation}
Later we will address this point elaborately.

%%%%%%%%%%%%%%%%%%%%%%%%%%%%
{\it Quantifying entanglement.}--
There are two well established quantities, negativity and concurrence, which fruitfully measure the entanglement between two quibts. Here we will find them corresponding to the state described by the density matrix (\ref{LTDM}). Any difference in them compare to those for (\ref{B2}) will signify the change in the entanglement between our two UD detectors.

 Negativity is defined as absolute value of the sum of negative eigenvalues of the partially transposed density matrix, derived from the PPT criterion \cite{Peres:1996dw}. The eigenvalues of the partially transposed matrix corresponding to (\ref{LTDM}) have been calculated in \cite{Chowdhury:2021ieg}. The analysis showed that out of four eigenvalues, the negative eigenvalue can be either $\lambda_{1}$ or $\lambda_{2}$, obtained as
\begin{equation}
\begin{aligned}
\lambda_{1,2}&\approx\frac{C^{2}}{2}\left(\sum_{j}(\alpha^{2}P_{j}+\gamma^{2}P_{j}'')+\alpha\gamma\text{Re}(P_{AB}'+\overline{P}_{AB}')\right)\\&\pm\left[\alpha\gamma-C^{2}\text{Re}\left(\alpha^{2}\zeta_{AB}+\gamma^{2}Y_{AB}+\alpha\gamma(M_{A}+M_{B})\right)\right]~.
 \end{aligned}
 \label{B5}
\end{equation}
The explicit forms of the different quantities are provided in Supplemental Material. For the present setup, the above reduces to
\begin{equation}
\lambda_{1,2} \approx \frac{C^{2}}{2}\sum_{j} \gamma^{2}P_{j}'' \pm \Big[\alpha\gamma - C^{2}\alpha\gamma \text{Re}(M_{A}+M_{B})\Big]~.
\label{B3}
\end{equation}
(A short discussion on this is elaborated in Supplemental Material as well).
Note that only $P''_j$ and Re($M_j$) are contributing and they are real and non-vanishing only when $\Delta E>mc^2$. Under this condition we have $(P''_j, \textrm{Re}(M_j))>0$.
Therefore when both $\alpha$ and $\gamma$ have same sign (e.g. Bell state of the form (\ref{B4}) with $\alpha=\gamma=1/\sqrt{2}$) then $\lambda_2$ is negative, while $\lambda_1$ is negative for opposite signs of $\alpha$ and $\gamma$ (e.g. Bell state of the form (\ref{B4}) with $\alpha=-\gamma=1/\sqrt{2}$). In both these situations the negativity is given by
\begin{equation}
\label{mNegativity}
\mathcal{N}=max\left\{0,\,|\alpha\gamma|-C^{2}(\gamma^2P''+2|\alpha\gamma|M)\right\}\,.
\end{equation}
Note that since $P''$ and $M$ are non-vanishing, even though the detectors are static, we will have less negativity. Therefore if an entangled pair of two-level detectors are kept at rest in the environment, there will be entanglement leakage and that phenomenon is driven by two effects -- spontaneous emission of individual detector and the vacuum fluctuation of the background field.

Another independent measure of entanglement is concurrence \cite{Wootters:1997id}. The importance of this quantity is due to its connection with the entanglement of formation. The concurrence is defined as
\begin{equation}\label{conc}
\mathscr{C}(\rho)=max\{0,\,\lambda'_{1}-\lambda'_{2}-\lambda'_{3}-\lambda'_{4}\}\,,
\end{equation}
where the $\lambda'$'s are the square-root of eigenvalues of the matrix $\rho_{AB}(\sigma_{y}\otimes\sigma_{y})\rho_{AB}^{\star}(\sigma_{y}\otimes\sigma_{y})$ and $\lambda'_{1}$ is the largest of them. For our density matrix (\ref{LTDM}), $\lambda'$'s are calculated in Supplemental Material.
For our model the concurrence is obtained as
\begin{equation}\label{conM}
\mathscr{C}(\rho) = max\left\{0,\,2|\alpha\gamma|-C^{2}\Big((|\alpha\gamma|+2\gamma^2)P''+2|\alpha\gamma|M\Big)\right\}\,.
\end{equation}
It decreases from the initial value. So again it is confirmed that there will be leakage of entanglement between the two UD detectors when they are not isolated from the environment and such is due to two phenomenon -- spontaneous emission of individual detector and vacuum fluctuation of quantum field.

Note that for this model both $P''$ and $M$ contain Dirac-delta function $\delta(0)$ and therefore are divergent. This is due to consideration of interaction for infinite time and such issue always arises naturally for the choice of switching function as unity. The same has also appeared in the original calculation for  transition probability of an accelerated detector. In this situation, making an analogy with the Fermi's golden rule, the transition probability per unit time (known as detector’s response function) is considered to be the relevant physical quantity (see for example section $3.3$ of  \cite{book:Birrell}). In the same sprit to tackle the present situation we define the change in negativity or concurrence per unit time as follows. Using the fact (\ref{B6})
one defines the change in negativity and concurrence per unit time as
\begin{equation}
\delta \dot{\mathcal{N}} = (\textrm{finite quantity}) \times \lim_{T\to\infty}\frac{1}{2\pi T} \int_{-T/2}^{T/2} du~,
\label{B7}
\end{equation}
and 
\begin{equation}
\delta \dot{\mathscr{C}} = (\textrm{finite quantity}) \times \lim_{T\to\infty}\frac{1}{2\pi T} \int_{-T/2}^{T/2} du~,
\label{B8}
\end{equation}
respectively, where the ``finite quantity'' in $\delta\dot{\mathcal{N}}$ is determined from $C^{2}(\gamma^2P''+2|\alpha\gamma|M)$ by removing the common factor $\delta(0)$ in it and so on.
In this case to make the perturbative calculation viable we satisfy the condition – initial negativity and concurrence per unit interaction time is greater than their rate of change. Then non-vanishing positive value of $\delta\dot{\mathcal{N}}$ or $\delta \dot{\mathscr{C}}$ can be regarded as the signature of degradation of initial entanglement (the same has been proposed earlier in \cite{Chowdhury:2021ieg} as well). Introduction of this idea of measuring the entanglement for the present model then shows the unavoidable leakage of initial entanglement which here depends on the energy gap $\Delta E$ of the detectors and mass of the scalar field $m$ under the condition $\Delta E>mc^2$. Moreover, the entanglement leakage will decrease with increasing mass ($m$) of the scaler field till $mc^2\sim\Delta E$. Also note that such is independent of the intra-distance ($|{\bf{d}}| = d$) between the detectors and hence situation remains same even if they sit together. Moreover no change will occur when $\Delta E = mc^2$. Interestingly, for massless field one finds degradation for all values of $\Delta E$. The latter discussion seems to indicate that when two entangled detectors are illuminated by photon, that will lead to decrease of quantum communication between them. Additionally it may also be noted that if only one detector (say, $A$) is switched on while other one (say, $B$) is shielded from environment, then also entanglement degradation will happen. But in this case $P''_B$ and $M_B$ will not contribute and hence degradation will be half of the earlier one.

{\it Discussion and implications.}-- Within this simple model we observed that two static entangled UD detectors looses communication when they are open to environment. This has been confirmed through negativity as well as concurrence of the two detectors system. Such phenomenon is driven by the spontaneous emissions of the individual detectors and the vacuum fluctuation of the background quantum fields. 
In this regard it may be mentioned that in literature it is already pointed out that in open quantum systems, the environment causes decoherence for the quantum systems (e.g see \cite{Zurek1991,Zurek2003}). Therefore, it is natural to expect that this may cause entanglement leakage (e.g. see \cite{2005PhR...415..207M}), which is observed here as well.
On the other hand in the detector context, it is well known that coupling with the background scaler field favours the generation of entanglement between two initially non-entangled detectors due to entanglement swapping from the field vacuum (see e.g. \cite{Reznik:2003mnx, Martin-Martinez:2015qwa, VerSteeg:2007xs, Kukita:2017etu, Cong:2020nec, Henderson:2018lcy,Barman:2021bbw, Henderson:2017yuv, Tjoa:2020eqh, Gallock-Yoshimura:2021yok, Barman:2021kwg, Pozas-Kerstjens:2015gta,Salton:2014jaa, Koga:2018the, Koga:2019fqh, Barman:2021bbw, Perche:2021clp}).
Interestingly here we observe the negative effect of vacuum fluctuation which was observed to be providing entanglement harvesting between two un-entangled UD detector.  
This degradation can be decreased by shielding one of the detectors from the environment.  Although the calculation has been done for eternal interaction between the detector and fields, but other types of switching function should not change the nature of the result. In the latter situation other terms in (\ref{B5}) will contribute, but in any case one can always find a negative eigenvalue and so negativity will decrease. For example, Gaussian type switching function yields non-vanishing value for $P_j$ and in that case the leakage will be driven by transition of detectors from ground state to excited state as well. Hence the phenomenon of entanglement leakage is quite unavoidable and therefore two entangled systems will suffer a spontaneous drainage of communication due to their surroundings.

%\textcolor{red}{In this regard it may be mentioned that the presence of dissipative terms in the evolution of the density matrix may lead to entanglement degradation \cite{2005PhR...415..207M}, where concurrence exponentially decays with the strength of the system-environment coupling. But in our case the time evolution of the density matrix has no dissipative contributions, i.e., the evolution is completely unitary ($d\rho/dt=i[H,\rho]$). Since the whole analysis has been done in a complete relativistic frame work, we feel that such phenomenon may be very natural in relativistic regime.}

Finally, we mention that the present model can have significant impact to understand more about black hole spacetimes. A Minkowski observer is equivalent to a freely-falling observer in black hole spacetime. Therefore the present result indicates that two initially entangled qubits' communication gets fade during their free-fall towards the horizon. Since the quantum nature of a black hole (particularly the black hole information paradox problem) is now being investigated in the light of quantum information, we feel that the present observation can be important in this field of study. This is not more than a suggestive one.

\vskip 3mm
\noindent
{\it Note added.}-- 
After completion of the draft, a very recent work \cite{Gallock-Yoshimura:2021xsy} in a similar motivation came to our notice. Our work is based on perturbative approach and therefore can be generalised to any type of switching function. While the previous one has been analysed non-perturbatively and therefore restricted to a specific switching function (Dirac-delta type). Therefore the results were very specific to the model used there. Moreover in the present discussion the actual causes from entanglement leakage have been illuminated.    

\vskip 3mm
\begin{acknowledgments}
{\it Acknowledgments.}-- DB would like to acknowledge Ministry of Education, Government of India for
providing financial support for his research via the PMRF May 2021 scheme. The research of BRM is 
partially supported by a START-UP RESEARCH GRANT (No. SG/PHY/P/BRM/01) from the Indian Institute of 
Technology Guwahati, India.

\end{acknowledgments}

\bibliographystyle{apsrev}

\bibliography{bibtexfile}

%%%%%%%%%%%%%%%%%%%%%%%%%%%
%\iffalse
%%%%%%%%%%%%%%%%%%%%%%%%%%%%%%%

\clearpage

%%%%%%%%%% Merge with supplemental materials %%%%%%%%%%
\pagebreak
\widetext
\begin{center}
\textbf{\huge{Supplemental Material}\\
\vskip 5mm
 \large Spontaneous entanglement leakage of two static entangled Unruh-DeWitt detectors}\\
Dipankar Barman, Angshuman Choudhury, Bhushan Kad and Bibhas Ranjan Majhi\\
{\it Department of Physics, Indian Institute of Technology Guwahati, Guwahati 781039, Assam, India.}
\end{center}

\email{dipankar1998@iitg.ac.in}
\email{c.angshuman@iitg.ac.in}
\email{k.bhushan@iitg.ac.in}
\email{bibhas.majhi@iitg.ac.in}

%%%%%%%%%% Merge with supplemental materials %%%%%%%%%%
%%%%%%%%%% Prefix a "S" to all equations, figures, tables and reset the counter %%%%%%%%%%
\setcounter{equation}{0}
\setcounter{figure}{0}
\setcounter{table}{0}
\setcounter{page}{1}
\makeatletter
\renewcommand{\theequation}{A\arabic{equation}}
\renewcommand{\thefigure}{S\arabic{figure}}
\renewcommand{\bibnumfmt}[1]{[S#1]}
\renewcommand{\citenumfont}[1]{S#1}
%%%%%%%%%% Prefix a "S" to all equations, figures, tables and reset the counter %%%%%%%%%%

\section{Section I: Elements of the time evolved density matrix Eq. (\ref{LTDM})}\label{Elements}
The explicit forms of the elements of the later time density matrix (\ref{LTDM}), upto second order perturbation expansion, are already calculated in \cite{BRM1}. Therefore, instead of giving a detailed derivation, here we just provide the general expressions of them. These are
\begin{equation}\label{rhoABelements}
\begin{aligned}
a_1 &=\gamma \gamma^{\star}- \gamma \gamma^{\star} C_{A}^{2}P_A^{\prime \prime}- \gamma \gamma^{\star} C_{B}^{2}P_B^{\prime \prime}- \alpha^{\star} \gamma C_{A}C_{B}\zeta_{A B}^{\star}-\alpha \gamma^{\star} C_{A}C_{B}\zeta_{A B}~; \\
a_2 &=\gamma \alpha^{\star}- \alpha \alpha^{\star} C_{A}C_{B}\zeta_{A B}- \gamma \gamma^{\star}  C_{A}C_{B}Y_{A B}^{\star}- \alpha^{\star} \gamma  C_{A}^{2}M_A- \alpha^{\star} \gamma C_{B}^{2} M_B~; \\
b_1 &= \alpha \alpha^{\star}  C_{A}^{2}P_A+ \gamma \gamma^{\star}  C_{B}^{2}P_B^{\prime \prime}+ \alpha^{\star} \gamma  C_{A}C_{B}P_{A B}^{\prime}+ \alpha \gamma^{\star}  C_{A}C_{B}P_{A B}^{\prime {\star}}~; \\
b_2 &=\alpha \gamma^{\star}  C_{A}^{2}\bar{P}_A^{\prime}+ \alpha^{\star} \alpha C_{A}C_{B} P_{A B}+ \gamma \gamma^{\star} C_{A}C_{B} X_{A B}^{\star}+ \gamma \alpha^{\star} C_{B}^{2} \bar{P}_B~; 
\\
c_1 &= \alpha^{\star} \gamma  C_{A}^{2}\bar{P}_A+ \alpha \gamma^{\star} C_{B}^{2} \bar{P}_B^{\prime}+ \gamma \gamma^{\star} C_{A}C_{B} X_{A B}+ \alpha \alpha^{\star} C_{A}C_{B}P_{A B}^{\star}~;\\
c_2 &= \gamma^{\star} \gamma  C_{A}^{2}P_A^{\prime \prime}+ \alpha \alpha^{\star} C_{B}^{2} P_B+ \gamma \alpha^{\star} C_{A}C_{B}\bar{P}_{A B}^{\prime}+ \alpha \gamma^{\star} C_{A}C_{B} \bar{P}_{A B}^{\prime {\star}}~; \\
d_1 &=\alpha \gamma^{\star}- \gamma^{\star} \alpha  C_{A}^{2}M_A^{\star}-\alpha \gamma^{\star}  C_{B}^{2}M_B^{\star}-\gamma \gamma^{\star} C_{A}C_{B}Y_{A B}- \alpha \alpha^{\star} C_{A}C_{B} \zeta_{A B}^{\star}~;\\
d_{2}&=\alpha\alpha^{\star} - \alpha \alpha^{\star} C_{A}^{2}P_A- \alpha \alpha^{\star} C_{B}^{2}P_B-C_{A}C_{B}\alpha^{\star}\gamma Y_{AB}-C_{A}C_{B}\alpha\gamma^{\star}Y_{AB}^{\star}
\,,
\end{aligned}
\end{equation}
where
\begin{equation}
\label{rhoABelementsF}
\begin{aligned}
P_{j}(\Delta{E})&=\int\int d\tau_{j}d\tau_{j}'\chi_{j}(\tau_{j})\chi_{j}(\tau_{j}')e^{i\Delta{E}(\tau_{j}-\tau_{j}')}G_{W}(x_{j}',x_{j})~;\\
P_{j}''(\Delta{E})&=\int\int d\tau_{j}d\tau_{j}'\chi_{j}(\tau_{j})\chi_{j}(\tau_{j}')e^{-i\Delta{E}(\tau_{j}-\tau_{j}')}G_{W}(x_{j}',x_{j})~;\\
\overline{P_{j}}(\Delta{E})&=\int\int d\tau_{j}d\tau_{j}'\chi_{j}(\tau_{j})\chi_{j}(\tau_{j}')e^{-i\Delta{E}(\tau_{j}+\tau_{j}')}G_{W}(x_{j}',x_{j})~;\\
\overline{P_{j}'}(\Delta{E})&=\int\int d\tau_{j}d\tau_{j}'\chi_{j}(\tau_{j})\chi_{j}(\tau_{j}')e^{i\Delta{E}(\tau_{j}+\tau_{j}')}G_{W}(x_{j}',x_{j})~;\\
M_{j}(\Delta{E})&=\int\int d\tau_{j}d\tau_{j}'\chi_{j}(\tau_{j})\chi_{j}(\tau_{j}')e^{i\Delta{E}(\tau_{j}-\tau_{j}')}\theta(\tau_{j}-\tau_{j}')\left(G_{W}(x_{j}',x_{j})+G_{W}(x_{j},x_{j}')\right)~;\\
P^{\star}_{AB}(\Delta{E})&=\int\int d\tau_{A}\tau_{B}'\chi_{A}(\tau_{A})\chi_{B}(\tau_{B}')e^{i\Delta{E}(\tau_{B}'-\tau_{A})}G_{W}(x_{A},x_{B}')~;\\
P'_{AB}(\Delta{E})&=\int\int d\tau_{A}\tau_{B}'\chi_{A}(\tau_{A})\chi_{B}(\tau_{B}')e^{-i\Delta{E}(\tau_{A}+\tau_{B}')}G_{W}(x_{A},x_{B}')~;\\
\overline{P'}_{AB}(\Delta{E})&=\int\int d\tau_{A}\tau_{B}'\chi_{A}(\tau_{A})\chi_{B}(\tau_{B}')e^{-i\Delta{E}(\tau_{A}+\tau_{B}')}G_{W}(x_{B}',x_{A})~;\\
X_{AB}(\Delta{E})&=\int\int d\tau_{A}d\tau_{B}'\chi_{A}(\tau_{A})\chi_{B}(\tau_{B}')e^{i\Delta{E}(\tau_{B}'-\tau_{A})}G_{W}(x_{B}',x_{A})~;\\
Y_{AB}(\Delta{E})&=\int\int d\tau_{A}d\tau_{B}'\chi_{A}(\tau_{A})\chi_{B}(\tau_{B}')e^{-i\Delta{E}(\tau_{B}'+\tau_{A})}\{iG_{F}(x_{A},x_{B}')\}~;\\
\xi_{AB}(\Delta{E})&=\int\int d\tau_{A}d\tau_{B}'\chi_{A}(\tau_{A})\chi_{B}(\tau_{B}')e^{i\Delta{E}(\tau_{B}'+\tau_{A})}\{iG_{F}(x_{A},x_{B}')\}\,.\\
\end{aligned}
\end{equation}
In the above $G_{W}(x,x') = \bra{0_M}\phi(x)\phi(x')\ket{0_M}$ is the positive frequency Wightman function. In Minkowski spacetime this is given by
\begin{equation}\label{Wightman}
G_{W}(x,x')=\frac{1}{16\pi^{3}}\int \frac{d^{3}p}{E_{\bf{p}}}e^{i\vec{p}\cdot(\vec{x}-\vec{x}')-iE_{\bf{p}}(t-t')}\,.
\end{equation}
Whereas $iG_{F}(x,x')$ is the Feynman propagator and it is related to $G_{W}(x,x')$ by the relation $iG_{F}(x_{A},x_{B}') = \theta(t-t')G_{W}(x,x') + \theta(t'-t)G_{W}(x',x)$.

For our choice of trajectories (see (\ref{trajectories})), $G_{W}(x,x')$ is time translational invariant with respect to detector's proper time. Using the coordinate transforms $T=\tau-\tau'$ and $\sigma=\tau-\tau'$ we can evaluate all the integrals in (\ref{rhoABelementsF}). The results are as follows (using infinite switching $\chi_{j}(\tau_{j})=1$):
\begin{eqnarray}
P_{j}(\Delta{E})&=&\int\int d\tau_{j}d\tau_{j}'e^{i\Delta{E}(\tau_{j}-\tau_{j}')}G_{W}(x_{j}',x_{j})
\nonumber
\\
&=&\frac{1}{16\pi^{3}}\int \frac{d^{3}p}{E_{\bf{p}}}\int\int\frac{dT\,d\sigma}{2}e^{i\Delta{E}\sigma}
e^{iE_{\bf{p}}\sigma}
\nonumber
\\
&=& \frac{1}{16\pi^{3}}\int \frac{d^{3}p}{E_{\bf{p}}}\int\frac{dT}{2}2\pi\delta(\Delta{E}+E_{\bf{p}})=0~;
\\
\nonumber
\\
P_{j}''(\Delta{E})&=&\int\int d\tau_{j}d\tau_{j}'e^{i\Delta{E}(\tau_{j}-\tau_{j}')}G_{W}(x_{j},x_{j}')
\nonumber
\\
&=& \frac{1}{16\pi^{3}}\int \frac{d^{3}p}{E_{\bf{p}}}\int\int\frac{dT\,d\sigma}{2}e^{i\Delta{E}\sigma}
e^{-iE_{\bf{p}}\sigma}
\nonumber
\\
&=& \frac{1}{16\pi^{3}}\int \frac{d^{3}p}{E_{\bf{p}}}\frac{1}{2}2\pi\delta(0)2\pi\delta(\Delta{E}-E_{\bf{p}})
\nonumber
\\
&=& \frac{\delta(0)}{8\pi}\int \frac{d^{3}p}{E_{\bf{p}}}\delta(\Delta{E}-E_{\bf{p}})
\nonumber
\\
&=& \frac{\delta(0)}{8\pi}\int_{0}^{\infty}\frac{4\pi p^{2}dp}{\sqrt{p^{2}c^{2}+m^{2}c^{4}}}\delta(\Delta{E}-\sqrt{p^{2}c^{2}+m^{2}c^{4}})
\nonumber
\\
&=& \frac{\delta(0)}{2}\int_{0}^{\infty}\frac{ p^{2}dp}{\sqrt{p^{2}c^{2}+m^{2}c^{4}}}\frac{\delta\left(p-\frac{1}{c}\sqrt{\Delta{E}^{2}-m^{2}c^{4}}\right)}{\frac{c\sqrt{\Delta{E}^{2}-m^{2}c^{4}}}{\Delta{E}}}
\nonumber
\\
&=& \frac{\delta(0)}{2}\frac{ \left(\frac{1}{c}\sqrt{\Delta{E}^{2}-m^{2}c^{4}}\right)^{2}}{\Delta{E}}\frac{1}{\frac{c\sqrt{\Delta{E}^{2}-m^{2}c^{4}}}{\Delta{E}}}
\nonumber
\\
&=& \frac{\delta(0)}{2c^{3}}\sqrt{\Delta{E}^{2}-m^{2}c^{4}}~;
\\
\overline{P_{j}}(\Delta{E})&=& \int\int d\tau_{j}d\tau_{j}'e^{-i\Delta{E}(\tau_{j}+\tau_{j}')}G_{W}(x_{j}',x_{j})
\nonumber
\\
&=& \frac{1}{16\pi^{3}}\int \frac{d^{3}p}{E_{\bf{p}}}\int\int\frac{dT\,d\sigma}{2}e^{-i\Delta{E}T}
e^{iE_{\bf{p}}\sigma}
\nonumber
\\
&=& \frac{1}{16\pi^{3}}\int \frac{d^{3}p}{E_{\bf{p}}}\frac{1}{2}2\pi\delta(\Delta{E})2\pi\delta(E_{\bf{p}})=0~;
\\
\nonumber
\\
\overline{P_{j}'}(\Delta{E})&=&\overline{P_{j}}(-\Delta{E})=0~;
\\
%\begin{eqnarray}
P^{\star}_{AB}(\Delta{E})&=&\int\int d\tau_{A}\tau_{B}'e^{i\Delta{E}(\tau_{B}'-\tau_{A})}G_{W}(x_{A},x_{B}')
\nonumber
\\
&=& \frac{1}{16\pi^{3}}\int \frac{d^{3}p}{E_{\bf{p}}}\int\int\frac{dT\,d\sigma}{2}e^{-i\Delta{E}\sigma}e^{-i\vec{p}\cdot\vec{d}-iE_{\bf{p}}\sigma}
\nonumber
\\
&=& \frac{1}{16\pi^{3}}\int \frac{d^{3}p}{E_{\bf{p}}}e^{-i\vec{p}\cdot\vec{d}}\frac{1}{2}2\pi\delta(0)2\pi\delta(\Delta{E}+E_{\bf{p}})=0~;
%\\
%\nonumber
\nonumber
\\
P'_{AB}(\Delta{E})&=&\int\int d\tau_{A}\tau_{B}'e^{-i\Delta{E}(\tau_{A}+\tau_{B}')}G_{W}(x_{A},x_{B}')
\nonumber
\\
&=& \frac{1}{16\pi^{3}}\int \frac{d^{3}p}{E_{\bf{p}}}\int\int\frac{dT\,d\sigma}{2}e^{-i\Delta{E}T}e^{-i\vec{p}\cdot\vec{d}-iE_{\bf{p}}\sigma}
\nonumber
\\
&=& \frac{1}{16\pi^{3}}\int \frac{d^{3}p}{E_{\bf{p}}}e^{-i\vec{p}\cdot\vec{d}}\frac{1}{2}2\pi\delta(\Delta{E})2\pi\delta(E_{\bf{p}})=0~;
\end{eqnarray}
%\\
%\nonumber
%\\
\begin{eqnarray}
M_{j}(\Delta{E})&=&\int\int d\tau_{j}d\tau_{j}'e^{i\Delta{E}(\tau_{j}-\tau_{j}')}\theta(\tau_{j}-\tau_{j}')\left(G_{W}(x_{j}',x_{j})+G_{W}(x_{j},x_{j}')\right)
\nonumber
\\
&=& \frac{1}{16\pi^{3}}\int \frac{d^{3}p}{E_{\bf{p}}}\int\int\frac{dT\,d\sigma}{2}e^{i\Delta{E}\sigma}\theta(\sigma)\left(e^{iE_{\bf{p}}\sigma}+e^{-iE_{\bf{p}}\sigma}\right)
\nonumber
\\
&=& \frac{1}{16\pi^{3}}\int  \frac{4\pi p^{2}dp}{E_{\bf{p}}}\frac{1}{2}\int_{-\infty}^{\infty}{dT}\int_{0}^{\infty}{d\sigma} \left(e^{i(\Delta{E}+E_{\bf{p}}+i\epsilon)\sigma}+e^{i(\Delta{E}-E_{\bf{p}}+i\epsilon)\sigma}\right)
\nonumber
\\
&=& \frac{\delta(0)}{4\pi}\int \frac{p^{2}\,dp}{E_{\bf{p}}}\,\left(PV\left(\frac{i}{\Delta{E}+E_{\bf{p}}}\right)+\pi\delta(\Delta{E}+E_{\bf{p}})+PV\left(\frac{i}{\Delta{E}-E_{\bf{p}}}\right)
+\pi\delta(\Delta{E}-E_{\bf{p}})\right)
\nonumber
\\
&=& \frac{\delta(0)}{4}\int\frac{p^{2}dp}{E_{\bf{p}}}\delta(\Delta{E}-E_{\bf{p}})+i\frac{\delta(0)}{4\pi}PV\int_{0}^{\infty} \frac{p^{2}\,dp}{E_{\bf{p}}}\,\left(\frac{2\Delta{E}}{\Delta{E}^{2}-E_{\bf{p}}^{2}}\right)
\nonumber
\\
&=& \frac{\delta(0)}{4c^{3}}\sqrt{\Delta{E}^{2}-m^{2}c^{4}}+i\frac{\delta(0)}{4\pi}PV\int_{0}^{\infty} \frac{p^{2}\,dp}{E_{\bf{p}}}\,\left(\frac{2\Delta{E}}{\Delta{E}^{2}-E_{\bf{p}}^{2}}\right)~;
%\end{eqnarray}
\\
\text{where} &PV(1/x)&\text{  represents the principal value of $1/x$.}
\nonumber\\
\nonumber
\\
\overline{P'}_{AB}(\Delta{E})&=&(P_{AB}'(-\Delta{E}))^{\star}=0~;
\\
\nonumber
\\
X_{AB}(\Delta{E})&=&\int\int d\tau_{A}d\tau_{B}'e^{i\Delta{E}(\tau_{B}'-\tau_{A})}G_{W}(x_{B}',x_{A})
\nonumber
\\
&=& \frac{1}{16\pi^{3}}\int \frac{d^{3}p}{E_{\bf{p}}}\int\int\frac{dT\,d\sigma}{2}e^{-i\Delta{E}\sigma}e^{i\vec{p}\cdot\vec{d}+iE_{\bf{p}}\sigma}
\nonumber
\\
&=& \frac{1}{16\pi^{3}}\int \frac{d^{3}p}{E_{\bf{p}}}e^{i\vec{p}\cdot\vec{d}}\frac{1}{2}2\pi\delta(0)2\pi\delta(\Delta{E}-E_{\bf{p}})
\nonumber
\\
&=& \frac{\delta(0)}{8\pi}\int \frac{d^{3}p}{E_{\bf{p}}}e^{i\vec{p}\cdot\vec{d}}\delta(\Delta{E}-E_{\bf{p}})
\nonumber
\\
&=& \frac{\delta(0)}{8\pi}\int \frac{d^{3}p}{E_{\bf{p}}}e^{i\vec{p}\cdot\vec{d}}\delta(\Delta{E}-E_{\bf{p}})
\nonumber
\\ 
&=& \frac{\delta(0)}{8\pi}\int_{0}^{\pi}\int_{0}^{\infty}\frac{2\pi\sin\theta d\theta p^{2}dp}{E_{\bf{p}}}e^{ipd\cos\theta}\delta(\Delta{E}-E_{\bf{p}})
\nonumber
\\ 
&=& \frac{\delta(0)}{4}\int_{0}^{\infty}\frac{p^{2}dp}{\sqrt{p^{2}c^{2}+m^{2}c^{4}}}\frac{2\sin(pd)}{pd}\delta(\Delta{E}-\sqrt{p^{2}c^{2}+m^{2}c^{4}})
\nonumber
\\
&=& \frac{\delta(0)}{2c^{3}}\sqrt{\Delta{E}^{2}-m^{2}c^{4}}~\frac{\sin\left(\frac{d}{c}\sqrt{\Delta{E}^{2}-m^{2}c^{4}}\right)}{\left(\frac{d}{c}\sqrt{\Delta{E}^{2}-m^{2}c^{4}}\right)}~;
\end{eqnarray}
%\\
%\nonumber
%\\
\begin{eqnarray}
Y_{AB}(\Delta{E})&=&\int\int d\tau_{A}d\tau_{B}'e^{-i\Delta{E}(\tau_{B}'+\tau_{A})}\{iG_{F}(x_{A},x_{B}')\}
\nonumber
\\
&=& \int\int d\tau_{A}d\tau_{B}'e^{-i\Delta{E}(\tau_{B}'+\tau_{A})}\{\theta(\tau_{B}'-\tau_{A})G_{W}(x_{B}',x_{A})
+\theta(\tau_{A}-\tau_{B}')G_{W}(x_{A},x_{B}')\}
\nonumber
\\
&=& \int\int d\tau_{A}d\tau_{B}'e^{-i\Delta{E}(\tau_{B}'+\tau_{A})}\{\frac{1}{16\pi^{2}} \int\frac{d^{3}p}{E_{\bf{p}}}
\nonumber
\\
&\times& [\theta(\tau_{B}'-\tau_{A})e^{i\vec{p}\cdot\vec{d}-iE_{\bf{p}}(\tau_{B}'-\tau_{A})}+\theta(\tau_{A}-\tau_{B}')e^{-i\vec{p}\cdot\vec{d}-iE_{\bf{p}}(\tau_{A}-\tau_{B}')}]
\}
\nonumber
\\
&=& \frac{1}{16\pi^{2}} \int\frac{d^{3}p}{E_{\bf{p}}} \int\int \frac{dTd\sigma}{2}e^{-i\Delta{E}T} [\theta(-\sigma)e^{i\vec{p}\cdot\vec{d}+iE_{\bf{p}}\sigma} +\theta(\sigma)e^{-i\vec{p}\cdot\vec{d}-iE_{\bf{p}}\sigma}]
\nonumber
\\
&=& \frac{1}{16\pi^{2}} \int\frac{d^{3}p}{E_{\bf{p}}} \int\frac{d\sigma}{2} [\theta(-\sigma)e^{i\vec{p}\cdot\vec{d}+iE_{\bf{p}}\sigma}+\theta(\sigma)e^{-i\vec{p}\cdot\vec{d}-iE_{\bf{p}}\sigma}]
2\pi\delta(\Delta{E})=0~;
\\
\nonumber
\\
\xi_{AB}(\Delta{E})&=&Y_{AB}(-\Delta{E})=0~.
\end{eqnarray}

\section{Section II: Negativity}\label{Negativity}
Partial transpose of the later time density matrix (\ref{LTDM}) is obtained as
\begin{equation}
\label{PTDM}
\rho^{T_{B}}_{AB}=\begin{pmatrix}
a_{1}&0&0&b_{2}\\
0&b_{1}&a_{2}&0\\
0&d_{1}&c_{2}&0\\
c_{1}&0&0&d_{2}\\
\end{pmatrix}\,.
\end{equation}
The eigenvalues of $\rho^{T_{B}}_{AB}$ for our static setup are 
\begin{equation}
\label{lambda12PTDM}
\begin{aligned}
&\lambda_{1,2}=\frac{1}{2}\left\{b_1+c_2\pm\sqrt{\left(b_1-c_2\right)^2+4 a_2 d_1}\right\} \\&=
\frac{1}{2}\left(\gamma^{2}(C_{A}^{2}P_{A}''+C_{B}^{2}P_{B}'')\pm\sqrt{\gamma^{2}(C_{A}^{2}P_{A}''-C_{B}^{2}P_{B}'')^{2}+4\alpha^{2}\gamma^{2}|1-C_{A}^{2}M_{A}-C_{B}^{2}M_{B}|^{2}}\right)
\\
&=\frac{1}{2}\left(\gamma^{2}(C_{A}^{2}P_{A}''+C_{B}^{2}P_{B}'')\pm\left(\gamma^{2}(C_{A}^{2}P_{A}''-C_{B}^{2}P_{B}'')^{2}+4\alpha^{2}\gamma^{2}(1-C_{A}^{2}M_{A}-C_{B}^{2}M_{B}-C_{A}^{2}M_{A}^{\star}-C_{B}^{2}M_{B}^{\star}\right.\right.\\&\left.\left.~~~~~~~~~~~~~~~~~~~~~~~~~~~~~~~~~~~~~~~~~~+C_{A}^{4}M_{A}M_{A}^{\star}+C_{B}^{4}M_{B}M_{B}^{\star}+C_{A}^{2}C_{B}^{2}M_{A}M_{B}^{\star}+C_{A}^{2}C_{B}^{2}M_{B}M_{A}^{\star})\right)^{1/2}\right);\\&=
\frac{1}{2}\left(\gamma^{2}(C_{A}^{2}P_{A}''+C_{B}^{2}P_{B}'')\pm2|\alpha\gamma|\left(1-C_{A}^{2}M_{A}-C_{B}^{2}M_{B}-C_{A}^{2}M_{A}^{\star}-C_{B}^{2}M_{B}^{\star}+\mathcal{O}(C^{4})\right)^{1/2}\right)\\&\simeq~
\frac{1}{2}\left(\gamma^{2}(C_{A}^{2}P_{A}''+C_{B}^{2}P_{B}'')\pm2|\alpha\gamma|\left(1-\frac{1}{2}(C_{A}^{2}M_{A}-C_{B}^{2}M_{B}-C_{A}^{2}M_{A}^{\star}-C_{B}^{2}M_{B}^{\star})\right)\right);\\&=
\frac{1}{2}\left(\gamma^{2}(C_{A}^{2}P_{A}''+C_{B}^{2}P_{B}'')\pm2|\alpha\gamma|\left(1-C_{A}^{2}\text{Re}\{M_{A}\}-C_{B}^{2}\text{Re}\{M_{B}\}\right)\right);
\end{aligned}
\end{equation}
and
\begin{equation}
\label{lambda34PTDM}
\begin{aligned}
&\lambda_{3,4}\\&=\frac{1}{2}\left\{a_1+d_2\pm\sqrt{\left(a_1-d_2\right)^2+4 b_2 c_1}\right\}\\&=\frac{1}{2}\left(\gamma^{2}(1-C_{A}^{2}P_{A}''-C_{B}^{2}P_{B}'')+\alpha^{2}\right.\\&\left.\pm\sqrt{(\gamma^{2}(1-C_{A}^{2}P_{A}''-C_{B}^{2}P_{B}'')-\alpha^{2})^{2}+4\gamma^{4}C^{2}_{A}C^{2}_{B}|X_{AB}
|^{2}} \right)
\\&=\frac{1}{2}\left(\gamma^{2}(1-C_{A}^{2}P_{A}''-C_{B}^{2}P_{B}'')+\alpha^{2}\pm
\sqrt{(\gamma^{2}-\alpha^{2})^{2}-2\gamma^{2}(\gamma^{2}-\alpha^{2})
\iffalse\textcolor{red}{c^{2}}\fi
(C_{A}^{2}P_{A}''+C_{B}^{2}P_{B}'')+\gamma^{4}\mathcal{O}(C^{4})}\right)\\&= \,
\frac{1}{2}\left(\gamma^{2}(1-C_{A}^{2}P_{A}''-C_{B}^{2}P_{B}'')+\alpha^{2}\pm(\gamma^{2}-\alpha^{2})
\sqrt{1-\frac{1}{(\gamma^{2}-\alpha^{2})}(2\gamma^{2}(C_{A}^{2}P_{A}''+C_{B}^{2}P_{B}'')+\gamma^{4}\mathcal{O}(C^{4}))}\right)\\&\simeq \,
\frac{1}{2}\left(\gamma^{2}(1-C_{A}^{2}P_{A}''-C_{B}^{2}P_{B}'')+\alpha^{2}\pm\left(
(\gamma^{2}-\alpha^{2})-\gamma^{2}(C_{A}^{2}P_{A}''+C_{B}^{2}P_{B}'')\right)\right)\\&=\gamma^{2}(1-C_{A}^{2}P_{A}''-C_{B}^{2}P_{B}'')~,~\alpha^{2}~
.
\end{aligned}
\end{equation}
Note that $\lambda_{3,4}$ can not be negative. While depending on the relative sign between $\alpha$ and $\gamma$, any one among $\lambda_{1,2}$ is negative.  Therefore the later ones are considered in (\ref{B3}).

\section{section III: Concurrence}\label{Concurrence}
Calculate first the following quantity:
\begin{equation}
\begin{aligned}
&\rho_{AB}(t)(\sigma_{y}\otimes\sigma_{y})\rho_{AB}^{\star}(t)(\sigma_{y}\otimes\sigma_{y})\\&=
\begin{pmatrix}
a_{1}&0&0&a_{2}\\
0&b_{1}&b_{2}&0\\
0&c_{1}&c_{2}&0\\
d_{1}&0&0&d_{2}\\
\end{pmatrix}
\begin{pmatrix}
0&0&0&-1\\0&0&1&0\\0&1&0&0\\-1&0&0&0
\end{pmatrix}\begin{pmatrix}
a_{1}&0&0&a_{2}^{\star}\\
0&b_{1}&b_{2}^{\star}&0\\
0&c_{1}^{\star}&c_{2}&0\\
d_{1}^{\star}&0&0&d_{2}\\
\end{pmatrix}\begin{pmatrix}
0&0&0&-1\\0&0&1&0\\0&1&0&0\\-1&0&0&0
\end{pmatrix}\\&=
\begin{pmatrix}
-a_{2}&0&0&-a_{1}\\0&b_{2}&b_{1}&0\\0&c_{2}&c_{1}&0\\-d_{2}&0&0&-d_{1}
\end{pmatrix}
\begin{pmatrix}
-a_{2}^{\star}&0&0&-a_{1}\\0&b_{2}^{\star}&b_{1}&0\\0&c_{2}&c_{1}^{\star}&0\\-d_{2}&0&0&-d_{1}^{\star}
\end{pmatrix}\\&=
\begin{pmatrix}
|a_{2}|^2 + a_{1} d_{2}& 0& 0& a_{1} a_{2} + a_{1} d_{1}^{\star}\\
 0&|b_{2}|^2 + b_{1} c_{2}& b_{1} b_{2} + b_{1} c_{1}^{\star}& 0\\
  0& b_{2}^{\star} c_{2} + c_{1} c_{2}& |c_{1}|^2 + b_{1} c_{2}& 0\\ 
  a_{2}^{\star} d_{2} + d_{1} d_{2}& 0& 0& |d_{1}|^2 + a_{1} d_{2}
\end{pmatrix}~.
\end{aligned}
\end{equation}
Then the square-root of the eigenvalues of the above matrix (till second order in $C$) are:
\begin{equation}\label{concEV12}
\begin{aligned}
\lambda'_{1,2}&=\sqrt{\frac{(|a_{2}|^{2}+|d_{1}|^{2}+2a_{1}d_{2})\pm\sqrt{(|a_{2}|^{2}-|d_{1}|^{2})^{2}+4a_{1}d_{2}(|d_{1}|^{2}+|a_{2}|^{2}+2\text{Re}\{a_{2}d_{1}\})}}{2}}\\
&=
\frac{1}{\sqrt{2}}\left(\gamma^{2}\alpha^{2}(1-C_{A}^{2}M_{A}-C_{B}^{2}M_{B})^{2}+\gamma^{2}\alpha^{2}(1-C_{A}^{2}M_{A}^{\star}-C_{B}^{2}M_{B}^{\star})^{2}+2\gamma^{2}\alpha^{2}(1-C_{A}^{2}P_{A}''-C_{B}^{2}P_{B}'')\right.\\&\left.\pm2\alpha\gamma\sqrt{1-C_{A}^{2}P_{A}''-C_{B}^{2}P_{B}''}\right.\\&\left.\times\sqrt{(\gamma^{2}\alpha^{2}(1-C_{A}^{2}M_{A}-C_{B}^{2}M_{B})^{2}+\gamma^{2}\alpha^{2}(1-C_{A}^{2}M_{A}^{\star}-C_{B}^{2}M_{B}^{\star})^{2}+2\alpha^{2}\gamma^{2}|1-C_{A}^{2}M_{A}-C_{B}^{2}M_{B}|^{2})}\right)^{1/2}~~~~
\\&=
\frac{|\gamma\alpha|}{\sqrt{2}}\left((4-2C_{A}^{2}M_{A}-2C_{B}^{2}M_{B}-2C_{A}^{2}M_{A}^{\star}-2C_{B}^{2}M_{B}^{\star}-2C_{A}^{2}P_{A}''-2C_{B}^{2}P_{B}''+\mathcal{O}_{1}(C^{4}))\right.\\&\left.~~~~~~\pm2\sqrt{1-C_{A}^{2}P_{A}''-C_{B}^{2}P_{B}''}\sqrt{4-8C_{A}^{2}\text{Re}\{M_{A}\}-8C_{B}^{2}\text{Re}\{M_{B}\}+\mathcal{O}_{2}(C^{4})}\right)^{1/2}~~~~\\&\approx
\frac{2|\gamma\alpha|}{\sqrt{2}}\left(\left(1-C_{A}^{2}\text{Re}\{M_{A}\}-C_{B}^{2}\text{Re}\{M_{B}\}-\frac{C_{A}^{2}P_{A}''+C_{B}^{2}P_{B}''}{2}\right)\right.\\&\left.~~~~~\pm\left(1-\frac{C_{A}^{2}P_{A}''+C_{B}^{2}P_{B}''}{2}\right)(1-C_{A}^{2}\text{Re}\{M_{A}\}-C_{B}^{2}\text{Re}\{M_{B}\})\right)^{1/2}~~~~\\&\approx
2|\gamma\alpha|\left(1-\frac{C_{A}^{2}\text{Re}\{M_{A}\}+C_{B}^{2}\text{Re}\{M_{B}\}}{2}-\frac{C_{A}^{2}P_{A}''+C_{B}^{2}P_{B}''}{4}\right)\,,~0\,.
\end{aligned}
\end{equation}

and
\begin{equation}\label{concEV34}
\begin{aligned}
\lambda'_{3,4}&=\sqrt{\frac{(|b_{2}|^{2}+|c_{1}|^{2}+2b_{1}c_{2})\pm\sqrt{(|b_{2}|^{2}-|c_{1}|^{2})^{2}+4b_{1}c_{2}(|b_{2}|^{2}+|c_{1}|^{2}+2\text{Re}\{b_{2}c_{1}\})}}{2}}\\
&=\frac{1}{\sqrt{2}}\left(
2\gamma^{4}C_{A}^{2}C_{B}^{2}|X_{AB}|^{2}+2\gamma^{4}C_{A}^{2}P_{A}''C_{B}^{2}P_{B}''\pm
2\gamma^{2}\sqrt{C_{A}^{2}P_{A}''C_{B}^{2}P_{B}''(2\gamma^{4}C_{A}^{2}C_{B}^{2}|X_{AB}|^{2}+2\gamma^{4}C_{A}^{2}C_{B}^{2}|X_{AB}|^{2}
)}\right)^{1/2}\\&=
C_{A}C_{B}\gamma^{2}\left(
|X_{AB}|^{2}+P_{A}''P_{B}''\pm2\sqrt{P_{A}''P_{B}''}|X_{AB}|
\right)^{1/2}\\&=\begin{cases}
C_{A}C_{B}\gamma^{2}\left(|X_{AB}|\pm\sqrt{P_{A}''P_{B}''}\right) & \text{~~~~~~~~~if,}~|X_{AB}|\geq\sqrt{P_{A}''P_{B}''}\\
C_{A}C_{B}\gamma^{2}\left(\sqrt{P_{A}''P_{B}''}\pm|X_{AB}|\right) &\text{~~~~~~~~~if,}~|X_{AB}|\leq\sqrt{P_{A}''P_{B}''}\end{cases}
\end{aligned}
\end{equation}
From (\ref{Expressions}), we always have $|X_{AB}|\leq\sqrt{P_{A}''P_{B}''}$ as $\frac{\sin x}{x}\leq1$. Therefore from above two sets we will consider the second set for $\lambda'_3$ and $\lambda'_4$. Using this then (\ref{conc}) will lead to (\ref{conM}).

%%%%%%%%%%%%%%%%%%%%%%%%%%%%%%%%

%\fi
%%%%%%%%%%%%%%%%%%%%%%%%

\end{document}